\documentclass[12pt]{article}

\usepackage{natbib}
\usepackage{amssymb}
\usepackage{amsmath}
\usepackage{amscd}
\usepackage{latexsym}
\usepackage{chemarrow} 
\usepackage[dvipdfm,pdfstartview=FitH,bookmarks=true,bookmarksopen=true,
colorlinks=true,citecolor=blue,,urlcolor=red]{hyperref}

\usepackage{abstract}

\def\bbar{\raisebox{4.pt}[0pt][0pt]{-}\hspace{-1.7mm}b}

\newcommand{\be}{\begin{equation}}
\newcommand{\ee}{\end{equation}}
\newcommand{\brr}{\begin{array}}
\newcommand{\ber}{\end{array}}
\newcommand{\bsf}{{\Psi({\bf x},{\bf q},t)}}
\newcommand{\bsfs}{{\left|\Psi({\bf x},{\bf q},t)\right|^2}}
\newcommand{\wfqt}{{\psi({\bf q},t)}}
\newcommand{\wfq}{{\Phi({\bf q})}}

\newcommand{\wfqo}{{\Phi(q)}}
\newcommand{\bfx}{{\bf x}}

\newcommand{\bfp}{{\bf p}}
\newcommand{\bfq}{{\bf q}}
\newcommand{\dfq}{{\rm d}{\bf q}}
\newcommand{\dfx}{{\rm d}{\bf x}}
\newcommand{\psii}{{\Phi_i}}
\newcommand{\hatf}{{\hat{F}}}
\newcommand{\hermf}{{\hat{F}^{\dag}}}

\newcommand{\mch}{\hat{\mathcal{H}}}
\newcommand{\mct}{\hat{\mathcal{T}}}
\newcommand{\mcv}{\mathcal{V}}

\newcommand{\idvs}{ideal individuals}

\newcommand{\hvi}{\harvarditem}

\begin{document}

\begin{center}
{{\Large {\bf A novel quantum theory of psychology}} \\ [5mm]
Jiao-Kai Chen } \\ [3mm]

{\it School of Physics and Information Engineering,\\ Shanxi Normal University, Linfen 041004, P. R. China \\ E-Mail: chenjk@sxnu.edu.cn, chenjkphy@outlook.com}
\end{center}

\vskip 1cm
\begin{abstract}
{The behavior coordinate system and the ideal individual model are presented. The behavior state of an ideal individual is assumed to be represented by a behavior state function. Based on
the ideal individual model, the behavior coordinate system and the quantum probability,
a novel quantum theory of psychology is offered here in a different way. It can
give some enlightening viewpoints through which some phenomena can be discussed from
a different perspective.}
\end{abstract}

\vskip 5mm


\section{Introduction}
Quantum mechanics is a fundamental theory in physics and is widespreadly applied to many areas, from natural sciences such as biology and chemistry \citep{Brookes17qb,Levine20qc} to social sciences such as economics and psychology \citep{Bagarello19b,Busemeyer12qcog,haven13sci,mansfield89qm,rossi94qp}.

Many phenomena in psychology which can not be explained by classical theories have been investigated by employing the quantum concepts and methods. In Ref. \citep{valle89qp}, discussions on the duality of human existence, i.e., the ``wave" side and ``particle" side of human existence is given analogous to the wave-particle duality in quantum physics.
In Ref. \citep{Busemeyer12qcog}, a human is considered to be in an indefinite state (formally called a superposition state) at each moment in time before a decision is made, and uncertainty is also discussed.
\cite{Khrennikov07intf} claim that quantum probability interference is present in cognition as well as in social phenomena.
In Refs. \citep{barros09inf,Busemeyer06gd}, the interference are discussed and the Schr\"{o}dinger equation is used.
In Ref. \citep{sozzo13}, a quantum probability model in Fock space is proposed.
In Ref. \citep{Bagarello18ah}, decision making is considered as decoherence and the operators of creation and annihilation are introduced. In Ref. \citep{Bagarello18he}, uncertainty in decision making is quantified with the aid of Heisenberg-Robertson inequality.
Introducing quantum principles to human judgement and decision making \citep{aerts94,atmanspacher02q,bordely98hu,busemeyerqps,khrennikov99cog} gives rise to a new field called quantum cognition in which puzzling behavioral phenomena are found and discussed by applying a probabilistic formulation with non-commutative algebraic principles \citep{Aerts09qcog,bruzaqcog,Busemeyer12qcog,khrennikovqb,Pothos13dec,wang13qcong,zwang2014qn}. The quantum concepts and methods are applied to psychology in a flood of literature. We only list a few, for example, see \citep{Aerts13qm,Bagarello19b,bruzaqcog,Busemeyer12qcog,haven13sci} and references therein.

We present a novel quantum theory of psychology in a different way. The old things can be recognized from a new point of view. In addition, there is always the hope that the new viewpoint will inspire an idea for the modification of present theories, a modification necessary to encompass present experiments.
In Sec. \ref{sec:behav}, the behavior coordinate system is proposed.
In Sec. \ref{sec:model}, a novel quantum theory of psychology is presented. We conclude in Sec. \ref{sec:conclusions}.

\section{Behavior coordinate system}\label{sec:behav}
In this section, the behavior is discussed and the behavior coordinate system is given. In this work, the case of human beings is discussed as an example, other species of organisms can be investigated similarly.
\subsection{Behavior}\label{subsec:behav}
As position of a point particle is the basic variable in Newtonian mechanics, behavior or behavior
position of an individual is the basic variable in psychology. Behavior refers to the observable actions of an individual. For human beings, behaviors are speech, body movement, emotional expression and so on.
Let $B$ be a behavior set, whose elements are the possible behaviors of all human beings. These possible behaviors are the collection of the behaviors of humans that have occurred in the past, are occurring at present and will occur in the future. Suppose $B_i$ is the set whose members are the possible behaviors of the $i$-th individual. $B_1$, $B_2$, $\cdots$ are the subsets of $B$, and there is
\be
B_1{\cup}B_2{\cup}\cdots=B.
\ee
In general, there are the relations
\be
B_i{\ne}B_j, \quad B_i{\cap}B_j{\ne}\O, \quad i,j=1,2,\cdots,
\ee
where $\O$ denotes the empty set.

An individual is not merely a capacity with all the content inserted by the environment,
\begin{equation}
C\autorightleftharpoons{F(x)}{}B,\quad x{\in}C,\; F(x){\in}B,
\end{equation}
where $F(x)$ is a relation which describes the physiological and psychological mechanisms and associates the set $C$ to the behavior set $B$. $C$ is the domain of $F$, which is a collection of all possible images in mind. $C$ is some like the mental reservoir proposed in \citep{Bagarello18ah}. $B$ is the range of $F$. The elements in $B$ can be realized while some elements in $C$ are only imagined in mind and can not be realized as behaviors.
It is expected that the elements in C can be described by quantum theory although they cannot be observed directly \citep{melkikh19con,woolf01qcon}, therefore, the elements in B which are mapped from C should be described also by quantum theory.

The interrelationship amongst $C$, $F(x)$ and $B$ is complex. In addition, $C$ and $F(x)$ will change rapidly or slowly with the growth of an individual because human characteristics are determined by some combination of genetic and environmental influences \citep{bouchard90}. Different $C$ and $F(x)$ result in different $B$ and then make species and individuals different.
Human differ radically from animal due to different $C$, $F(x)$ and $B$. Human is rational and is constrained by law, culture, rite, etc.. For animal, the relation from $C$ to $B$ will be more direct and relatively simpler. Therefore, animals will be the better research objects for investigating the quantum phenomena in psychology.

\subsection{Behavior coordinate system}

Except for the intrinsic quantities such as age and gender, psychological observables, such as motivation, emotion and personality are supposed to be functions of behavior and the derivative of behavior with respect
to time. In other words, these quantities can be revealed by behavior and its derivative

Let the behavior coordinate system be a coordinate system that specifies each behavior
point uniquely in a behavior space by a set of numerical behavior coordinates \citep{hock15ft,rosenhan73ax}. The reference lines are $q_1$-axis, $q_2$-axis, $q_3$-axis and so on. To describe the behaviors of human beings, we can assume that $q_1$-axis is the speech-axis, $q_2$-axis is the body-movement-axis and $q_3$-axis is the emotional-expression-axis.

For simplicity, we assume temporarily the behavior coordinate system is a Cartesian coordinate system, and the behavior coordinate space is a n-dimensional Euclidean space $\mathbb{R}^n$. Although the behavior coordinate system is far from being well established now, it is useful for solving many problems and it can give some enlightening results.

\section{Quantum theory of psychology}\label{sec:model}
In this section, a novel quantum theory of psychology is presented. Some quantities have been named in psychology, for example, volition is regarded as a vectorlike quantity \citep{valle89qp}. For simplicity, we borrow nomenclature in physics to denominate these quantities.

\subsection{Ideal individual}

An ideal individual is an idealization of humans or other species of organisms. We propose an assumption that for an ideal individual every possible behavior can occur with equal probability. These possible behaviors are elements in the behavior set $B$. Different behaviors show different properties and make distinction among different species and different individuals.

\subsection{Behavior state function and probability interpretation}
The behavior state of an ideal individual is assumed to be represented by a behavior state function $\bsf$ which represents one behavior pattern, where $t$ is time, $\bfx$ denotes the spatial space vector $\bfx=(x,y,z)$. The spatial position in psychology is some different from that in physics because an organism has volume and cannot be regarded as a particle in psychology although it can be regarded as a particle in physics which is defined to be an object of insignificant size. $\bfq$ denotes the behavior space vector $\bfq=(q_1,q_2,q_3,\cdots)$. For human beings, $q_1$ is the speech coordinate, $q_2$ is the body-movement coordinate, $q_3$ is the emotional-expression coordinate, and $\cdots$ denotes other coordinates.

We assume that the superposition principle holds for $\Psi$. If $\psii$ describe possible states of an ideal individual,
\be
\Psi=\sum_{i=1}^{n}c_i\psii
\ee
is also a possible state. The superposition principle demands that the dynamics equation for $\Psi$ should be a linear equation.

The choice of probability models \citep{khrennikov99cog,Pothos09qp,Pothos13dec,sozzo13} is crucial. In general, there is the probability $P$,
\be
P=|\Psi|^{\alpha},\quad \alpha>0.
\ee
If $\alpha=1$ and $\Psi$ is a real function, $P$ is the classical probability and there is no interference effects although the superposition principle holds \citep{feynman66qm}. As $\alpha{\ne}1$, the problems will be discussed in the $L^{\alpha}$ space and the interference occurs. If $\alpha=2$, $P$ becomes a quantum probability according to the Born's hypothesis which reproduces the interference effects, and $\Psi$ is a vector in the Hilbert space which is a $L^2$ space.

Adopting the Born hypothesis which gives the probability interpretation of the wave function in quantum mechanics, the behavior state function $\bsf$ is a probability amplitude and $\bsfs$ is interpreted as the probability of finding an ideal individual at a behavior point $\bfq$ at time $t$ and spatial space $\bfx$. There is the normalization condition
\be
\int\bsfs \dfq\dfx=1.
\ee
The integration is over whole behavior space and whole spatial space.

In physics, the position of a particle and its derivative respective to time are used to discuss mechanical problems. Similarly, in psychology, the behavior of an ideal individual and its derivative respective to time are concentrated.
In many cases, the spatial coordinates have small or even no effects on an individual's behavior and then the spatial coordinates $\bfx$ can be neglected. In some cases, the spatial space can be taken as part of the environment. It is not spatial space but the environment that affects behaviors greatly. In consequence, the behavior coordinates are highlighted in these cases and the spatial coordinates can be integrated out
\be
\wfqt=c\int{\bsf}{\dfx},\quad \int|\wfqt|^2 \dfq=1,
\ee
where $c$ is a normalization factor. The superposition principle holds also for $\wfqt$. In many references \citep{Busemeyer12qcog,haven13sci}, the behavior state function $\wfqt$ has been used actually to discuss problems in different forms although the behavior variable $\bfq$ is not given explicitly.

\subsection{Operators}

Similar to quantum mechanics, we assume that each measurable quantity $F$ has a corresponding operator $\hatf$.
Any quantity that can be measured is an observable. It is obvious that the result of a measurement of a dynamical variable must always be a real number \citep{dirac58qm}, therefore, the operator should be Hermitian, $\hatf=\hermf$. The superposition principle leads to the linearity of the operators. Thus the operators should be linear and Hermitian. The eigenvalue equation for the operator $\hatf$ reads
\be
\hatf\phi=f\phi,
\ee
where $\phi$ is an eigenfunction of $\hatf$ with eigenvalue $f$.

The behavior position operator is the behavior space vector itself
\be\label{opq}
\hat{\bfq}=\bfq.
\ee
Its components are
\be\label{opq2}
\hat{q}_1=q_1,\;\;\hat{q}_2=q_2,\;\;\hat{q}_3=q_3,\;\;\cdots.
\ee
The behavior momentum operator is assumed to be expressed as
\be
\hat{\bfp}=-{\rm i}{\bbar}\nabla,
\ee
and its components are
\be\label{opp}
\hat{p}_1=-{\rm i}\bbar\frac{\partial}{{\partial}q_1},\;\;
\hat{p}_2=-{\rm i}\bbar\frac{\partial}{{\partial}q_2},\;\;
\hat{p}_3=-{\rm i}\bbar\frac{\partial}{{\partial}q_3},\;\;\cdots,
\ee
where ${\rm i}=\sqrt{-1}$. $\bbar$ is a new constant which plays the same role in quantum psychology as $\hbar$ plays in quantum physics \citep{Busemeyer11qp,khrennikov99cog}. Using Eqs. (\ref{opq2}) and (\ref{opp}), the commutators are obtained
\begin{eqnarray}\label{possbra}
[\hat{q}_i,\hat{p}_j]=\hat{q}_i\hat{p}_j-\hat{p}_j\hat{q}_i={\rm i}\delta_{ij}\bbar,
\quad [\hat{q}_i,\hat{q}_j]=0, 
\quad [\hat{p}_i,\hat{p}_j]=0,
\end{eqnarray}
where $\delta_{ij}$ is the Kronecker delta function.

\subsection{Time evolution}

Suppose the time evolution equation of a behavior state function can be written in the form of the Schr\"{o}dinger equation as \citep{khrennikov99cog,Pothos13dec,Triffet96qcons}
\be\label{schd}
{\rm i}{\bbar}\frac{\partial\wfqt}{{\partial}t}=\mch\wfqt,
\ee
where $\mch$ is the Hamiltonian operator. The concrete form of $\mch$ is unknown to us now due to its complexity and the unwell-established behavior coordinate system. The differential equation (\ref{schd}) and the Schr\"{o}dinger equation in quantum physics take the same form, however, they are fundamentally different. In Ref. \citep{Busemeyer06gd,Pothos09qp}, the Schr\"{o}dinger equation is used to discuss quantum dynamics of human decision-making, but the new key constant $\bbar$ is not given.

Suppose an ideal individual is in a behavior state $\psi(\bfq, t_0)$ at $t_0$ and in behavior state $\psi(\bfq, t)$ at $t$. These two behavior states are related by an operator which is called the time-evolution operator
\be\label{schwf}
\psi(\bfq, t)=\hat{U}(t,t_0)\psi(\bfq, t_0).
\ee
The time-evolution operator has the properties
\be
\hat{U}(t_0,t_0)=1,\quad \hat{U}^{\dag}(t,t_0)\hat{U}(t,t_0)=1,\quad
\hat{U}(t_2,t_0)=\hat{U}(t_2,t_1)\hat{U}(t_1,t_0),\;\;(t_2{\ge}t_1{\ge}t_0).
\ee
Using Eqs. (\ref{schd}) and (\ref{schwf}), there is \citep{Busemeyer06gd,khrennikov99cog,Pothos09qp,Pothos13dec}
\be\label{sche}
{\rm i}{\bbar}\frac{\partial}{{\partial}t}\hat{U}(t,t_0)=\mch\hat{U}(t,t_0).
\ee
From Eq. (\ref{sche}), we have
\be
\hat{U}(t,t_0)=\exp\left[-\frac{{\rm i}}{\bbar}{\mch}(t-t_0)\right].
\ee
Using the time-evolution operator, we have
\be\label{heisf}
\hat{F}_H(t)=\hat{U}^{-1}\hat{F}\hat{U},
\ee
where $\hat{U}=\hat{U}(t,t_0)$. By differentiation of Eq. (\ref{heisf}), we obtain
\be
\frac{\partial}{{\partial}t}\hat{F}_H=-\frac{{\rm i}}{\bbar}[\hat{F}_H,\mch].
\ee
It is called the Heisenberg's equation for the operator $\hat{F}_H$, which is radically different from the Heisenberg's equation in quantum physics.

$\mch$ will be in a complicated form and is expected to take the following form
\be
\mch=\mct+\mcv(V),
\ee
where $\mct$ is related to the properties of an individual, $V$ is the external environment which is independent on the individual and $\mcv(V)$ is the mapped environment by the individual from the external environment $V$ and his mind. Consequently, $\mcv(V)$ may be different for different individuals for a given $V$. In some cases, $\mcv(V){\approx}V$. In some cases, $\mcv(V)=V+V_c$, where $V_c$ is a correction term. Sometimes, $\mcv(V)$ will be in a complicated form.

Let the Hamiltonian $\mch$ be not explicitly time-dependent. The variables $\bfq$ and $t$ of the time-dependent differential equation can be separated. With $\wfqt={\wfq}f(t)$ we have two differential equations,
\be\label{ssche}
{\rm i}\bbar\frac{{\partial}f}{{\partial}t}=\mathcal{E} f(t),\quad {\mch}\wfq=\mathcal{E}\wfq.
\ee
Solving the first equation in the above equation gives the time factor $f(t)=\exp[-{\rm i}{\mathcal{E}}t/\bbar]$. The second equation in Eq. (\ref{ssche}) is a stationary differential equation.

\subsection{Uncertainty principle}

Let two observables be described by operators $\hat{A}$ and $\hat{B}$. The commutator of these two operators is written as
\be
[\hat{A},\hat{B}]=\hat{A}\hat{B}-\hat{B}\hat{A}={\rm i}\hat{C}.
\ee
There is the Heisenberg's uncertainty principle \citep{greiner02qm}
\be\label{huncert}
\overline{({\Delta}A)^2}\hspace{1mm}\overline{({\Delta}B)^2}\ge\frac{{(\overline{C})^2}}{4},
\ee
where
\begin{align}
&\overline{A}=\int\psi^{\ast}\hat{A}{\psi}{\rm d}{\bf q},\;
\overline{B}=\int\psi^{\ast}\hat{B}{\psi}{\rm d}{\bf q},\;
\overline{C}=\int\psi^{\ast}\hat{C}{\psi}{\rm d}{\bf q},\;\nonumber\\
&{\Delta}\hat{A}=\hat{A}-\overline{A},\;\; {\Delta}\hat{B}=\hat{B}-\overline{B},\nonumber\\
&\overline{({\Delta}A)^2}=\overline{A^2}-\overline{A}^2,\;\;
\overline{({\Delta}B)^2}=\overline{B^2}-\overline{B}^2.
\end{align}
Eq. (\ref{huncert}) can be written in a short form as
\be\label{huncert2}
{\Delta}A{\Delta}B\ge\frac{1}{2}|{\overline{C}}|,
\ee
where ${\Delta}A=\sqrt{\overline{A^2}-\overline{A}^2}$, ${\Delta}B=\sqrt{\overline{B^2}-\overline{B}^2}$. From Eqs. (\ref{possbra}) and (\ref{huncert2}), we have
\be
{\Delta}q_i{\Delta}p_i\ge\frac{\bbar}{2},\quad i=1,2,\cdots.
\ee

\subsection{Identical ideal individuals}
Two ideal individuals are identical if there should be no experiment that detects any intrinsic
difference between them. The identical ideal individuals are those individuals that have the
same age, gender, etc. and behave in the same manner under equal conditions.
If the difference between the spatial positions and some intrinsic properties of individuals are
indistinguishable approximately or can be neglected to some extent as the ideal individuals
are congregated with high population density, these ideal individuals are assumed to be
identical.
In Ref. \citep{chen19ii}, we propose the identical ideal individual hypothesis. According to this hypothesis, the identical ideal individuals should be classified into two classes: the bosonic individuals and the fermionic individuals. The bosonic individuals can occupy the same behavior state while the fermionic individuals can not be in the same behavior state. We conjecture that the bosonic species will be small in number or be very different from the fermionic species if they exist in nature.
We propose that human beings and many species of animals are fermionic, which can not occupy the same behavior state according to the Pauli exclusion principle.

The existence of the personal space \citep{pshall66,pskatz37,pssommer59,psstern38,psuexhull37} and the behavior differentiation under high population density condition \citep{pdcalhoun62,pdEvans79pd,pdMarsden72pd} are two important and confusing issues
in psychology. The natures of these two phenomena remain mysterious and irrelated in old
theories. In Ref. \citep{chen19ii}, these two phenomena are explained theoretically in an unified approach by using the identical {\idvs} hypothesis. The existence of the personal space is a quantum effect in spatial space caused by the identity of the {\idvs} while the behavior differentiation under high population density condition is a quantum effect in behavior space.

\subsection{The simplest toy model}\label{sec:simpmodel}
Let
\be
\bfp=m\frac{d\bfq}{dt}=m{\bf v},
\ee
where $\bfp$ is the behavior momentum, $m=\bfp/{\bf v}$ is referred to as the behavior inertial mass, ${\bf v}$ is the behavior velocity. Due to the complexity of the discussed problems, we assume temporarily that the behavior kinetic energy takes the simple form
\be
\mathcal{T}=\frac{\bfp^2}{2m}.
\ee
In case of one dimension, the stationary differential equation (\ref{ssche}) is written as
\be\label{scheo}
\mathcal{E}\wfqo=\mch\wfqo,\quad \mch=\frac{\hat{p}^2}{2m}+\mcv(V),\quad \hat{p}=-{\rm i}\bbar\frac{\partial}{{\partial}q},
\ee
where $\mathcal{E}$ is the behavior energy.

According to Eqs. (\ref{possbra}) and (\ref{huncert2}), we have the Heisenberg's uncertainty in one dimension
\be
{\Delta}q{\Delta}p{\sim}\bbar.
\ee
In general, it is impossible to determine a human's motivation according to one action. The motivation is often confined by observing enough behaviors. Suppose the motivation can be expressed in a function of behavior and behavior momentum, $F(q,p)$. If $\hat{F}(q,\hat{p})=\underset{m,n}{\sum}c_{mn}q^m\hat{p}^n$ where $c_{mn}$ are coefficients, there is
\be\label{commf}
[q,\hat{F}]={\rm i}{\bbar}\frac{\partial \hat{F}}{\partial \hat{p}}, \quad
[\hat{p},\hat{F}]=-{\rm i}{\bbar}\frac{\partial \hat{F}}{\partial q}.
\ee
Then, we can obtain the uncertainty relation from Eqs. (\ref{huncert2}) and (\ref{commf})
\be
{\Delta}q{\Delta}F\ge\frac{1}{2}|{\overline{C}}|, \quad \hat{C}={\bbar}\frac{\partial \hat{F}}{\partial \hat{p}}.
\ee

\section{Conclusions}\label{sec:conclusions}

We propose the behavior coordinate system and the ideal individual model. Base on the the behavior coordinate system and the ideal individual model, the behavior state of an ideal individual is assumed to be represented by a behavior state function and then a novel quantum theory of psychology is proposed. Although the behavior coordinate system is far from being well established, this quantum theory of psychology can give some enlightening viewpoints through which some phenomena can be discussed from a different perspective. In fact, many problems have been discussed qualitatively or even quantitatively by employing the behavior state function.

\section*{Acknowledgment}
The author would like to thank Prof. Fabio Bagarello for helpful comments, and Prof. Evgeny Novikov for useful suggestions.


\end{document}